\documentclass[12pt,preprint]{aastex}

\shorttitle{HST and Keck Laser on $z\sim0.7$ Galaxy}
\shortauthors{Steinbring et al.}

\input epsf
\def\plotone#1{\centering \leavevmode
\epsfxsize=1.0\columnwidth \epsfbox{#1}}
\def\plottwo#1#2{\centering \leavevmode
\epsfxsize=0.5\columnwidth \epsfbox{#1}
\epsfxsize=0.5\columnwidth \epsfbox{#2}}

\begin{document}

\title{{\small CATS: Optical to Near-Infrared Colors of the Bulge and Disk of Two $z=0.7$ Galaxies Using HST and Keck Laser Adaptive Optics Imaging}\altaffilmark{1}}

\author{E. Steinbring\altaffilmark{2}, J. Melbourne\altaffilmark{3}, A. J. Metevier\altaffilmark{3,4}, D. C. Koo\altaffilmark{3}, M. R. Chun\altaffilmark{5}, L. Simard\altaffilmark{2}, J. E. Larkin\altaffilmark{6}, \& C. E. Max\altaffilmark{3}}

\altaffiltext{1}{All authors except L.S. are affiliated with the Center for Adaptive Optics.}
\altaffiltext{2}{Herzberg Institute of Astrophysics, National Research Council Canada, 
Victoria, BC V9E 2E7, Canada}
\altaffiltext{3}{UCO/Lick Observatory, Department of Astronomy and Astrophysics, 
University of California, Santa Cruz, CA 95064}
\altaffiltext{4}{NSF Astronomy \& Astrophysics Postdoctoral Fellow}
\altaffiltext{5}{Institute for Astronomy, University of Hawaii, 640 North A\'ohoku Place, Hilo, HI 96720}
\altaffiltext{6}{Division of Astronomy \& Astrophysics, University of California, Los Angeles, CA 90095}

\begin{abstract}
We have employed laser guide star (LGS) adaptive optics (AO) on the Keck II telescope to obtain near-infrared (NIR) images in the Extended Groth Strip (EGS) deep galaxy survey field. This is a continuation of our Center for Adaptive Optics Treasury Survey (CATS) program of targeting $0.5<z<1$ galaxies where existing images with the {\it Hubble Space Telescope} (HST) are already in hand.  Our AO field has already been imaged by the Advanced Camera for Surveys (ACS) and the Near Infared Camera and Multiobject Spectrograph (NICMOS).  Our AO images at 2.2$\mu$m ($K'$) are comparable in depth to those from HST, have Strehl ratios up to 0.4, and FWHM resolutions superior to that from NICMOS.  By sampling the field with the LGS at different positions, we obtain better quality AO images than with an immovable natural guide star.  As examples of the power of adding LGS AO to HST data we study the optical to NIR colors and color gradients of the bulge and disk of two galaxies in the field with $z=0.7$.
\end{abstract}


\section{Introduction}\label{introduction}

Although the bulges of disk-dominated galaxies can appear to be the remnants of long-ago mergers, their kinematics may point to close ties with ongoing formation processes in the disk \citep[for reviews, see][and references therein]{Kormendy2004, Wyse1997}. The strong correlation of bulge and disk color seen in local galaxies \citep{deJong1996, Peletier1996} suggests a shared history, with similar ages and metallicities in their stellar populations. So it is important to probe bulge and disk colors at significant look-back times.  Of particular interest might be the link between bulge and central black hole mass, for example, as that correlation plausibly develops at an epoch when bulges are first undergoing collapse or mergers, providing fuel for the black hole \citep{Maggorian1998, Ferrarese2000, Gebhardt2000}.

Separating the bulge from disk in images of 
high-redshift field galaxies is a difficult task though, because bulges are small; typically less than a few kpc across in
the local universe \citep{Andredakis1995}. At redshift $z=0.5$ each kpc corresponds to only 0\farcs16 in a 
concordance cosmology ($H_0=70$ km ${\rm s}^{-1}$ Mpc, $\Omega_{\rm
m}=0.3$, $\Omega_{\Lambda}=0.7$) which we adopt throughout. The {\it Hubble Space Telescope} (HST) has been used to decompose many $z>0.5$ field galaxies into bulges and disks in the optical \citep{Schade1995, Simard2002, Koo2005}. But equally high resolution near-infrared (NIR) color maps would also be valuable, as they should lessen contamination by dust and probe older stellar populations, especially for galaxies with $z>0.7$ where the $V$ filter desirably samples rest-frame $U$, which is below the 4000~\AA~break. Our initial efforts to do high-resolution NIR photometry of $0.5<z<1$ galaxies with natural-guide-star (NGS) AO were hampered by poor control over the point-spread function (PSF) \citep{Steinbring2004}. Better results are now possible with the new generation of laser guide-star (LGS) AO systems.

The motivation for LGS AO is to increase sky coverage by providing a pointable artificial beacon on which to guide.
LGS AO will also limit anisoplanatism, and thus improve image quality, by allowing the wavefront sensor and deformable mirror to be aligned 
with the target, instead of a bright nearby 
star. One complication is that the laser spot can only be used to sense high-order aberrations, not phase tip and tilt. 
Thus, a nearby star, which can be several magnitudes fainter than that used in full NGS mode, is still needed 
to account for image motion. A separate sensor guides on this ``tip-tilt star" and drives a fast steering mirror.  Despite this 
added complexity, tests with 
the Lick LGS AO system show that the resulting point-spread function (PSF) is no harder to
constrain than with a conventional NGS system \citep{Steinbring2005}.

Here we discuss our ongoing effort
to employ the latest AO technology to obtain images of high-redshift field galaxies. This work is part of the 
Center for Adaptive Optics (CfAO) Treasury Survey (CATS) \citep{Melbourne2005, Koo2006}.
We have employed Keck LGS AO in the
Extended Groth Strip (EGS), an HST field extensively imaged in both the optical and near infrared.

The data are presented in Section~\ref{data}, including a description of the improved PSF provided by the laser and the comparison of AO and HST images. Details of photometry and image decomposition follow for two $z=0.7$ galaxies in the field in Section~\ref{photometry}, and color gradients between the bulge and disk are discussed further in Section~\ref{discussion}.

\section{Data and Reductions}\label{data}

The AO field was chosen from the EGS where HST data exist from several programs. Besides WFPC2 data in F606W and
F814W (GO-5090: PI-Groth, GO-8698: PI-Mould), ACS Wide Field Camera (WFC) data in the equivalent filters and NICMOS NIC3 in F160W have been taken (GO-10134: PI-Davis).  We make use of the ACS data in our analysis, as these are deeper and have finer pixel sampling than the previous WFPC2 data. Exposures were 2260 s in ACS F606W, 2100 s in ACS F814W, and 640 s in NICMOS F160W, enough to provide excellent $S/N$ ($>10$) in the peak pixel (0\farcs05 for ACS and 0\farcs20 for NICMOS) for galaxies brighter than $I\sim22$ mag. All HST images were processed with the standard pipelines. The reduced ACS data were kindly provided to us by
 J. Lotz, and the reduced NICMOS data by S. Kassin. 
Fluxes were converted to Vega magnitudes using the zeropoints in the Data Handbooks for ACS \citep{Pavlovsky2006} and NICMOS \citep{Mobasher2004}. 
We will henceforth refer to the HST filters F606W, F814W, and F160W as $V$, $I$, and $H$, respectively.

The EGS ACS data were taken in a rectangular mosaic designed to avoid bright stars. Thus, no $V<12$ mag star is available to serve as a natural guide star (NGS) for the Keck AO system.  Several stars in EGS are, however, bright enough ($V<18$) for tip-tilt guiding in the LGS mode, and one of the stars happens to lie within one of the parallel NICMOS pointings. Using this $V=14$ star to guide tip-tilt correction, we obtained Keck II LGS-mode AO observations with the NIRC2 camera with the $K'$ filter on 2 March 2005. See \cite{Wizinowich2006} and \cite{LeMignant2006} for more information on the Keck II LGS AO system and its operation.  Individual exposures were 120 s, obtained in a nonrepeating circular dither pattern of 4\arcsec~radius, for a total integration of 5640 s. For these small dithers no re-acquisition is necessary with each laser move, and so overheads are no more than for NGS mode. The center of the dither pattern was aligned with the center of the NICMOS field, which required an 18\arcsec~offset between the laser spot and tip-tilt star. Exposures were obtained with NIRC2 with 0\farcs04 pixel wide-field mode (40\arcsec~$\times$~40\arcsec) which slightly undersamples the Keck diffraction limit in the NIR (${\rm FWHM}\approx0\farcs05$ at 1.6$\mu$m).  This is a good match to NICMOS NIC3 field (50\arcsec~$\times$~50\arcsec) although the 0\farcs20 NIC3 pixels significantly undersample the HST PSF (${\rm FWHM}\approx0\farcs16$ at 1.6$\mu$m). Data reductions closely followed those described in our earlier NGS work \citep{Steinbring2004}.
A photometric zeropoint of $Z_{K'}=24.84\pm0.07$ (from the Keck LGS AO NIRC2 webpages) was used to convert to standard Johnson $K'$.

Apart from the tip-tilt guide star, we detect seven objects with Keck AO in the region of overlapping HST coverage. One is a star, bright but unsaturated, which is suitable as a PSF estimator for HST.  Three are galaxies of known redshift, listed in Table~\ref{table_targets}. Two of these were measured as part of the DEEP2 redshift survey \citep{Davis2007}. A third has a photometric redshift obtained from the Canada-France-Hawaii Telescope Legacy Survey \citep{Ilbert2006}. There is another galaxy in the ACS images of unknown, although presumably low, redshift. It is only marginally detected at $K'$.  No redshifts are published for the remaining two objects either, and they are too faint in all bands for our analysis. 

\subsection{HST Image Quality}\label{image_quality}

The HST PSF is well characterized. The field distortions of ACS and NICMOS are modest and can be modeled in detail with the Tiny Tim software \citep{Krist1995}.  The PSF Strehl ratio - the peak intensity relative to that of the ideal diffraction pattern - is high, approaching $S=0.86$ for NICMOS NIC3 $H$ at optimal focus \citep{Barker2007}. So to first order the FWHM will describe the PSF, and differences in Strehl ratio can indicate PSF variation over the field.  

The HST PSF FWHM was measured on the bright but unsaturated star in the field using a Gaussian fit with the IRAF task IMEXAM.  The ACS WFC has a PSF FWHM of 0\farcs11 in $V$ and 0\farcs10 in $I$. The NICMOS NIC3 $H$ PSF has ${\rm FWHM}\approx0\farcs35$. Variation in PSF FWHM for the ACS data was estimated by comparing the PSF star in the field with another suitable star 74\arcsec~to the southwest, outside the overlapping NICMOS coverage. This was found to differ in FWHM by less than 0\farcs01 ($\Delta {\rm FWHM} = 4$\% in $V$, 6\% in $I$). We cannot measure PSF variation in the NICMOS field as there is only one unsaturated star available, but variation in PSF FWHM over the NIC3 field is also known to be slight, less than 0\farcs02 ($\Delta {\rm FWHM}\sim6$\%; Barker et al., 2007). Variation in the PSF Strehl ratio for our ACS data can be estimated from the relative flux-normalized maxima in the IMEXAM fits. This is a difference of $\Delta S=0.02$ in $V$ and $\Delta S=0.09$ in $I$ over a range of 74\arcsec.  The largest separation of PSF and target is 37\arcsec, so variation in Strehl ratio over our field will be less.  The case should be similar for NICMOS $H$, as its PSF FWHM variation is comparable to ACS.

\subsection{Temporal and Spatial Variations in AO Imaging}\label{AO_data}

Variations in LGS AO performance, both in time and angular separation from the guide source, are similar to those experienced in NGS mode. See \cite{vanDam2006a} and \cite{vanDam2006b} for a discussion of Keck II LGS AO performance characterization. Changes in seeing with time result in fluctuations in AO correction over the entire field of view. Increasing angular separation $\theta$ between the target and laser spot results in a drop in Strehl ratio proportional to $\exp{\theta}^{5/3}$.  It also increases radial elongation of the PSF towards the laser spot.
Although we did not record the seeing at $V$, previous experience with Keck AO indicates that for the correction achieved it was $\sim$0\farcs7, and fairly stable during the night. 
Two steps were taken to track changes in AO performance and characterize the data.  First, we interleaved the scientific observations with short, unsaturated exposures of the tip-tilt star. This establishes the PSF at the tip-tilt guide location immediately prior to each scientific exposure. Second, we followed the scientific observations by duplicating the dither pattern used for science observations in a dense star field. The globular cluster M5 was observed, which was at comparable airmass to EGS and is familiar from our previous AO calibration work \citep{Steinbring2002}. We used a $V=12$ star for tip-tilt guiding.  That this was brighter than the star used for EGS should not impact the results, as $V=14$ is already sufficient for excellent tip-tilt correction. We applied the same offset as the science observations and duplicated the circular dither pattern. We then shut off the laser, and continued the observations in M5 using this tip-tilt star as the NGS-mode guide for comparison with our LGS observations.  

\begin{figure}
\plotone{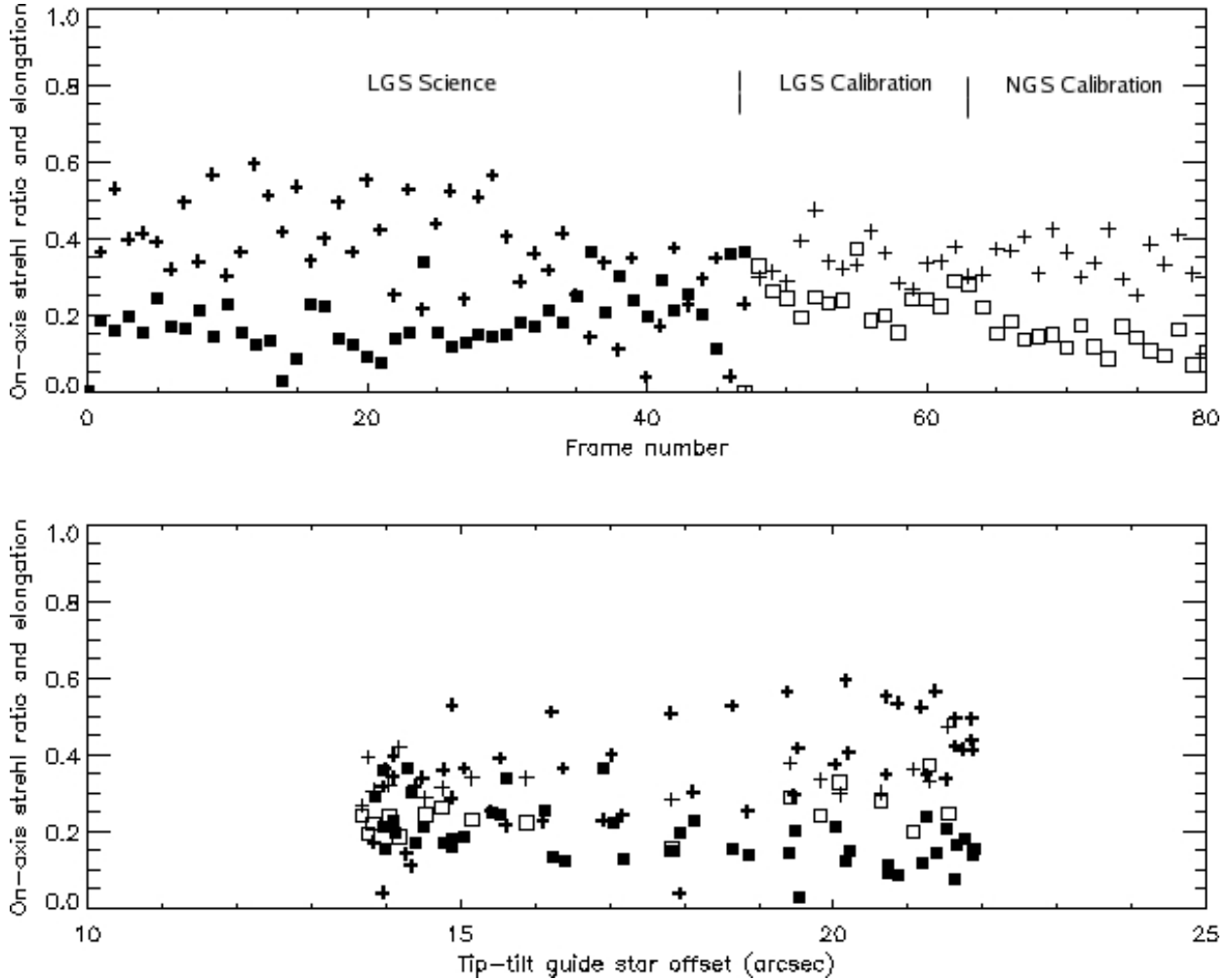}
\caption{Top panel shows extrapolated on-axis Strehl ratio (squares) and elongation (crosses) for each frame we observed. Science frames are indicated by filled symbols and thick crosses; calibration frames by open symbols and thin crosses. Note that approximately two minutes separates each frame for the science data, with less than one minute between each calibration frame. The bottom panel plots the LGS results as a function of separation between the tip-tilt guide star and the laser spot.}
\label{plot_strehl_ratio}
\end{figure}

These calibration data were combined to give the frame-by-frame performance of the AO system. The Strehl ratio of the unsaturated images of the tip-tilt star and another faint star visible in the scientific exposures were used to interpolate the ``on-axis" Strehl ratio - the value at the center of each frame, coincident with the laser spot - via an $\exp{\theta}^{5/3}$ fit.  This was repeated
in each of the M5 fields.  In the latter case, the fit was to all Strehl ratios of isolated unsaturated stars.
Measured Strehl ratios are plotted in Figure~\ref{plot_strehl_ratio}, along with the 
elongation of the PSF at the tip-tilt guide star position.
Image quality improved during the observations and peaked at a Strehl ratio of $S=0.38$ just before we moved the telescope to M5. These last science images also correspond to the least elongated images, and the strong anti-correlation with Strehl ratio is mirrored in the calibration data. Strehl ratio and elongation data are plotted again in the lower panel of Figure~\ref{plot_strehl_ratio} as a function of separation between the tip-tilt guide star and the laser spot. Note the expected trend towards increasing PSF elongation with increasing offset.  However, one can see that Strehl-ratio performance is fairly stable as a function of dither position, and so it is probably reasonable to neglect the effect of anisoplanatism due to separation of the tip-tilt star and the laser spot.

\begin{figure}
\plotone{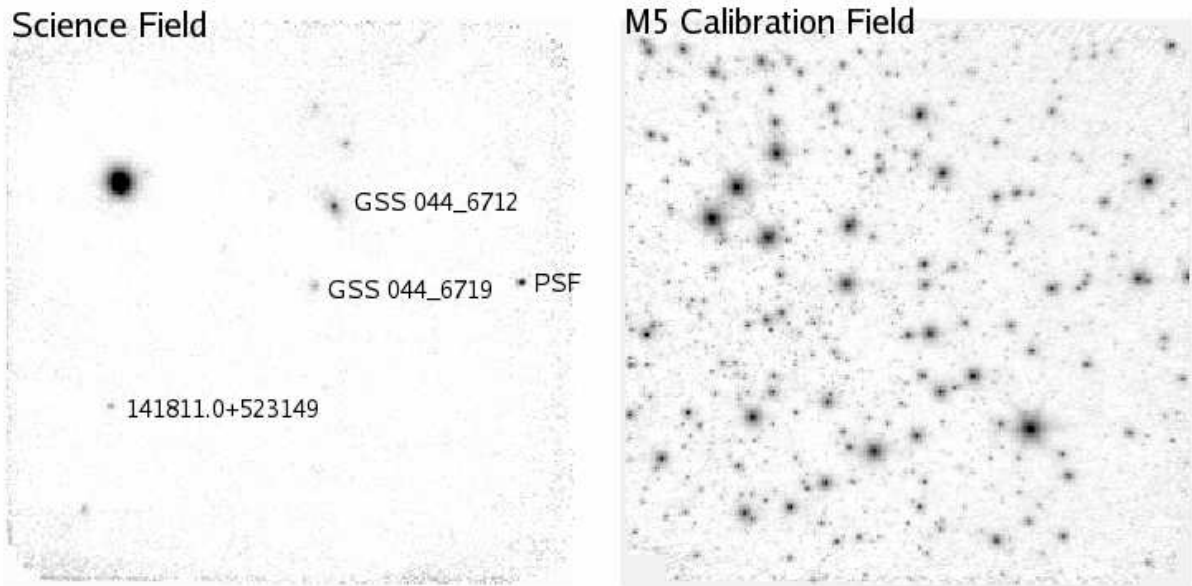}
\caption{Keck LGS AO NIRC2 $K^{'}$ image of the EGS field (left) and M5 PSF calibrator field (right). North is up, and east is to the left in these 49\arcsec~$\times$~49\arcsec~fields. Tip-tilt guiding was provided with the bright star at upper-left in the EGS field, and the corresponding central one of three bright stars in the M5 field (indicated by `GS').  The latter was also used for full NGS-mode correction (image not shown). We reproduced the laser dither pattern of the scientific observations in the calibration field.  This provides surrogate PSF stars throughout the field.  The star used as a PSF estimator in the science field is indicated by `PSF'.}
\label{figure_compare}
\end{figure}

The final images of both the target and the M5 field were combined after pruning poor data. Image quality and per-pixel $S/N$ were
found to be optimal by setting a cutoff for on-axis Strehl ratio of $S=0.15$.  This eliminated 10 of the EGS frames or 
about a fifth of the data; and similarly for M5.  The Strehl ratio at the center of the combined M5 frames is $S=0.23$, which is consistent with the interpolated mean on-axis Strehl ratio for the science frames ($S=0.25$, see Figure~\ref{plot_strehl_ratio}). It is worth emphasizing that this is based on a model fit to the on-axis Strehl ratio in each frame, not a measured value there.  Although the true PSF field variation may differ from the model, this method allows us to uniformly compare all of the M5 and EGS frames based on the best correction expected in each. The results are shown in Figure~\ref{figure_compare}. The variation in image quality over the 
combined images can be determined from the M5 field.
Although LGS observations still suffer from anisoplanatism, the image quality across the
field should be improved over our previous NGS observations. This is due to being able to move
 the laser spot around the field, 
and thus {\it spread out the best correction over the combined frames}.

\subsection{Laser vs Natural Guide Star AO Performance}\label{lgs_vs_ngs}

\begin{figure}
\plotone{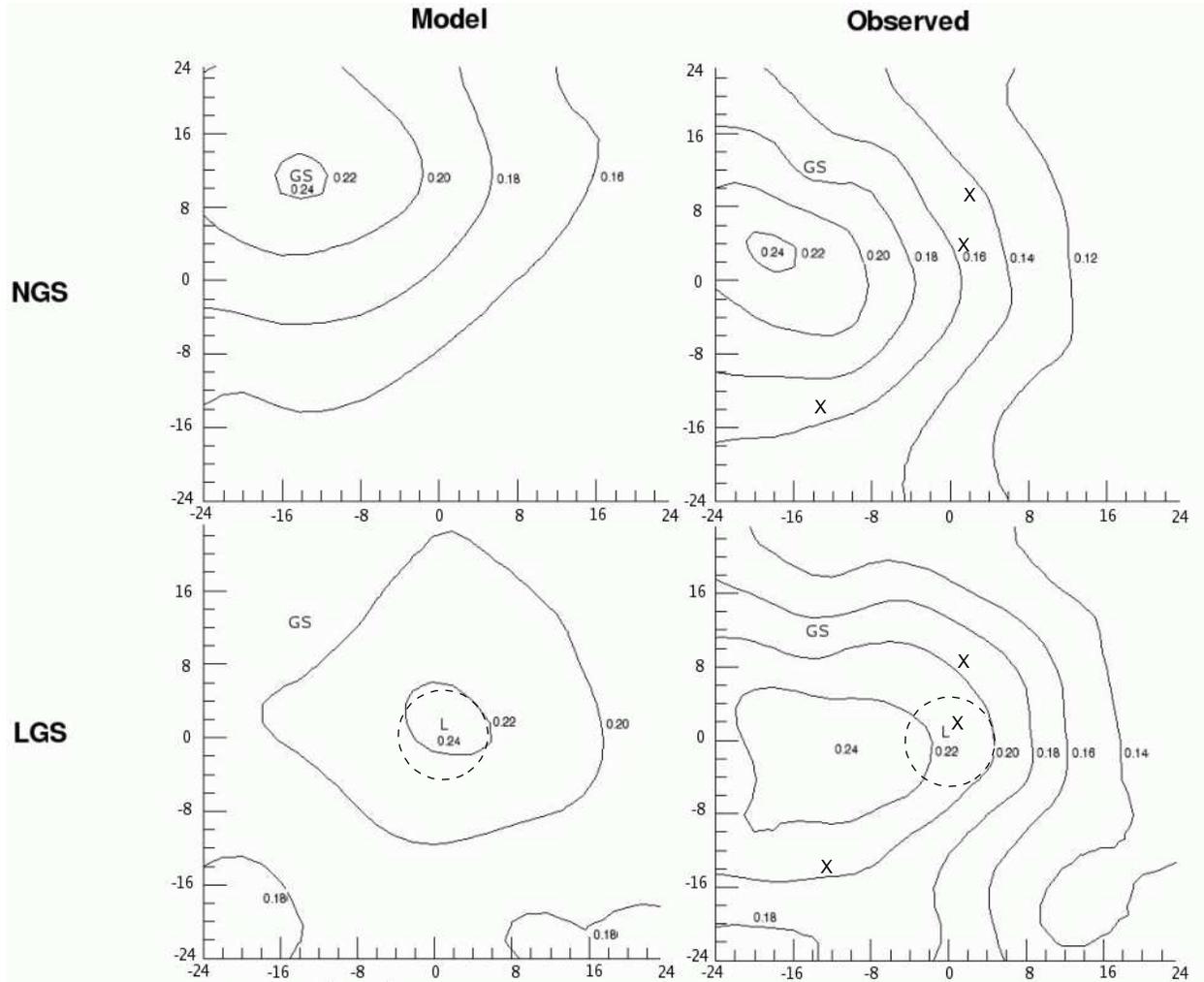}
\caption{Contours of equal Strehl ratio over the NIRC2 field for NGS mode (top row) and LGS mode (bottom row). The location of the NGS is marked by `GS' and a dashed ring labelled `L' indicates the positions of the laser spot. Axis labels are units of arcseconds from the center of the field.  The left hand panels are model results. Laser
observations are expected to have less spatial variation in the PSF. The best correction should also be achieved in a more useful location - offset from the guide star, close to the center of the field.  Although the models do not account for field distortion in the camera optics, they give a qualitatively correct picture. We observed a significant improvement with LGS performance over NGS, and less variation in the PSF between target locations, each indicated by an 'X'.}
\label{figure_contours}
\end{figure}

To help characterize the improved performance of LGS over NGS we employed an
analytic model of AO performance discussed in \cite{Steinbring2005}. A synthetic starfield
was generated by producing a spatially varying PSF for the location of every star in each M5 frame.
This is a simple PSF model comprising a diffraction-limited Gaussian core and a Gaussian halo into which core light is scattered as the Strehl ratio diminishes. Anisoplanatic degradation of the Strehl
ratio is determined by the fit to the calibration data. These fake frames were then combined and their
Strehl ratio determined for each isolated star in the same manner as the original data.  The results are shown along the left hand column in Figure~\ref{figure_contours}. These are contour plots of equal Strehl ratio. Notice how the region of best correction in the LGS observation is predicted to be larger than for NGS and closer to the center of the field. This is due to dithering the laser spot in a broad circle around the center of the field, which is indicated by the dashed curve. The observed results are shown in the right-hand panels. 
The advantage of this LGS geometry is demonstrated by inferring the Strehl ratio at the positions of the three targets in the field.  The greatest disparity in Strehl ratio between any two target positions is roughly $\Delta S\approx0.17-0.14=0.03$ for NGS.  For LGS, correction is both better and more uniform, with a maximum disparity of $\Delta S\approx0.21-0.19=0.02$. There is clearly some spatial dependence not accounted for in the model, but this could be due to, for example, static aberrations in the NIRC2 camera. Overall, the model seems to give a qualitatively correct picture of LGS performance, which indicates that visiting many locations in the field with the laser is the right approach to reduce PSF spatial variation.

\subsection{Combined Dataset}

\begin{figure}
\plotone{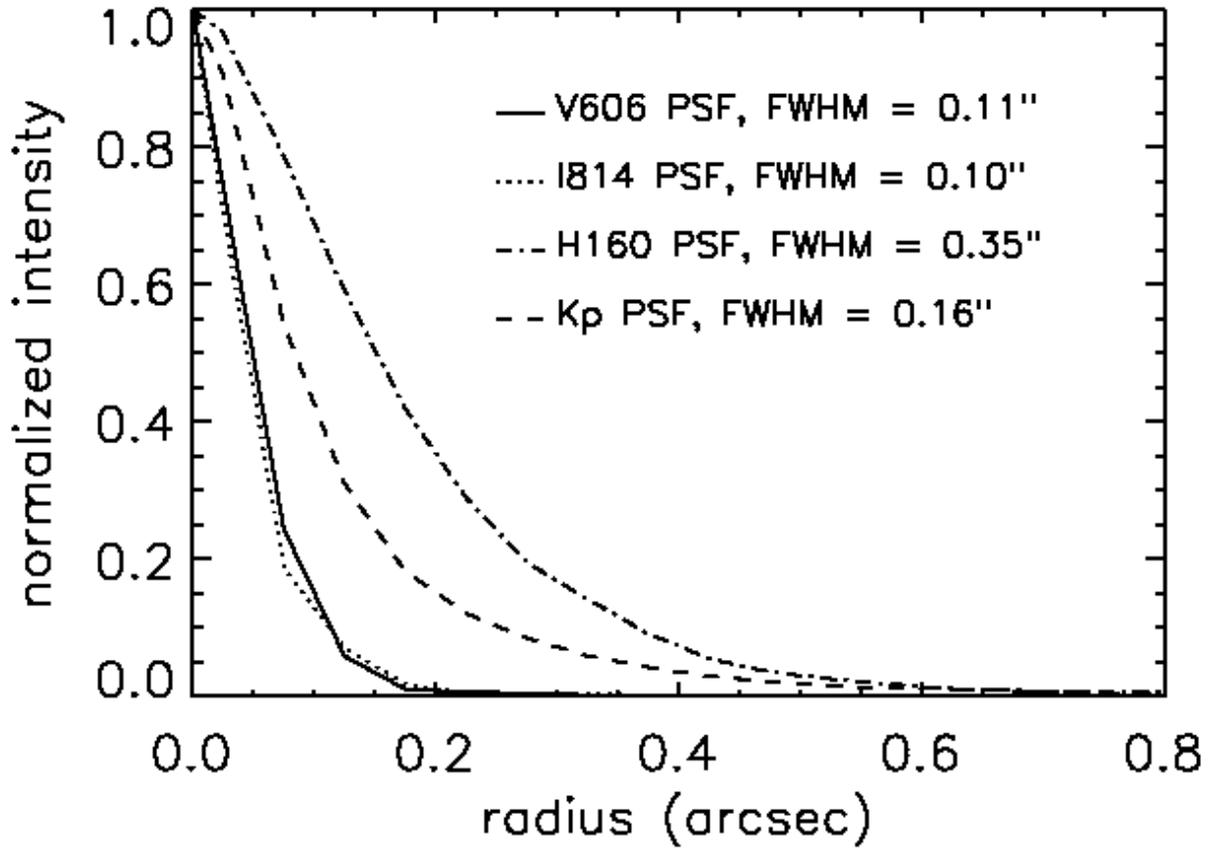}
\caption{Keck LGS AO $K^{'}$ PSF after pruning data with Strehl ratio less than 0.15. The HST ACS and NICMOS PSFs are shown for comparison.}
\label{figure_psf_plot}
\end{figure}

As a final reduction step, the combined NIRC2 science and M5 calibration frames were resampled to match the ACS pixellation.  No smoothing was done prior to resampling. Figure~\ref{figure_psf_plot} shows a radial plot of the PSF star in the science frames. This has a FWHM of 0\farcs16, measured with a Gaussian using the IRAF task IMEXAM.  Although this is a radial average, the elongation of the PSF is slight, less than 0.2.  The PSF FWHM is smaller at the positions of the three targets: for GSS 044\_6719 it is 0\farcs11; GSS 044\_6712, 0\farcs14; and 141811.0+523149, 0\farcs15. The ACS and NICMOS profiles are also plotted for comparison.

\begin{figure}
\plotone{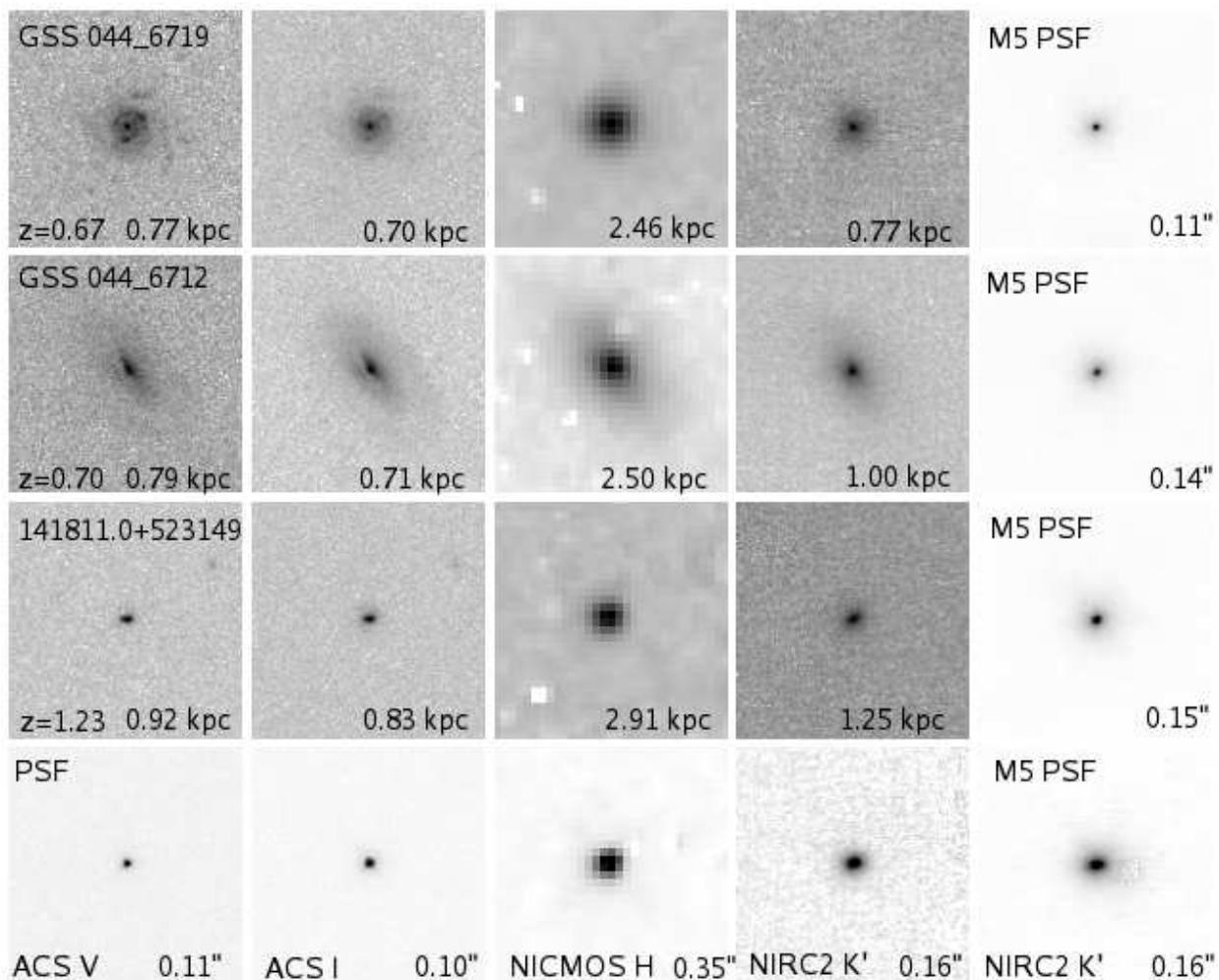}
\caption{Images of the three targets and PSF star in each filter. North is up, east is to the left. The field of fiew of each panel is 6\arcsec~$\times$~6\arcsec. The surrogate AO PSFs from the M5 calibration frame are shown along the right-hand column. Note the good agreement between the $K'$ PSF from the science field and that obtained from M5. The physical scale corresponding to the PSF FWHM is given for each image. For $z=0.7$ the central wavelengths of $V$, $I$, $H$, and $K'$ are 0.32, 0.48, 0.97, and 1.29 $\mu$m respectively in the restframe of the galaxy (0.24, 0.35, 0.72, and 0.95 $\mu$m for $z=1.3$).} 
\label{figure_panels}
\end{figure}

The three targets are shown
in Figure~\ref{figure_panels} along with the HST data. Each panel is 6\arcsec$\times$6\arcsec. 
North is up, and east is left.  The PSF star is shown along the bottom row in all bands with the same scaling. There is a good match between this PSF in $K'$ and that obtained from M5, both having $\rm {FWHM}=$~0\farcs16. Galaxy 141811.0+523149 ($z=1.228$) seems somewhat elongated in ACS $V$ and $I$, but it is only marginally resolved in our NIRC2 $K'$ image and no disk is evident. We therefore choose to exclude it from further analysis and instead focus on the two $z=0.7$ targets, both of which are resolved in all bands.

\subsection{GSS 044\_6719}\label{egs_e}

\begin{figure}
\plottwo{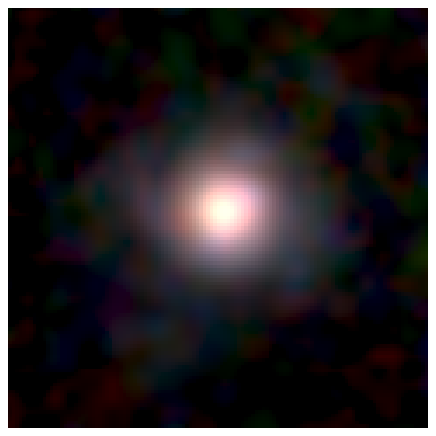}{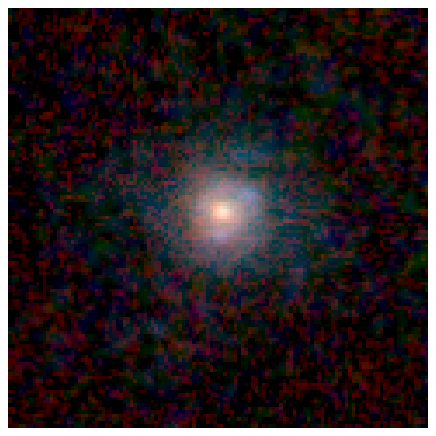}
\caption{Color images of GSS 044\_6719, our best-resolved galaxy, based on ACS $V$, $I$, and NICMOS $H$ data (top) and ACS $V$, $I$, and NIRC2 $K'$ data (bottom). ACS images have been degraded to match the NICMOS $H$ PSF. Note how the improved resolution of NIRC2 helps reveal the bulge. The field of view and orientation of each panel is the same as Figure~\ref{figure_panels}.}
\label{figure_color}
\end{figure}

The ACS $V$ and $I$ images reveal this $z=0.67$ galaxy to be nearly face on, with the disk easily seen in
both images. Also apparent are well defined spiral arms with
bright star-formation regions.  The central bulge-like core is very compact; only a few pixels across in
these images, and appears more pronounced in the $I$ image. With 0\farcs20 pixels and a PSF FWHM of 0\farcs35, the NICMOS $H$ image does not resolve the central core. In contrast, our $K'$ galaxy image has a PSF more comparable to that of the ACS and shows a marginally resolved core with clear signs of the disk. Figure~\ref{figure_color} shows color images produced by combining the ACS $V$ and $I$ with either NICMOS $H$ or NIRC2 $K'$. In each case the ACS images have been degraded with a Gaussian filter to match the resolution of either the NICMOS or NIRC2 PSF. Notice that the improved resolution of NIRC2 over NICMOS NIC3 helps reveal the separate bulge and disk components of the galaxy.

\subsection{GSS 044\_6712}\label{egs_c}

The ACS $V$ and $I$ images of this $z=0.70$ galaxy are dramatic, clearly showing a smooth, inclined disk and an elongated core, more bulge-like in $I$. There is a broad dust lane towards the center of the disk in the $I$ image. The NICMOS $H$ image also shows the galaxy disk and a core which is elongated in the same direction as the disk. The clump of white pixels
just north of center in this image is an artifact due to a poorly removed cosmic ray.  The disk is also faintly
visible in our NIRC2 $K'$ image. The $K'$ core is asymmetric, elongated in the same sense as in ACS $I$.

\section{Photometry}\label{photometry}

Synthetic aperture photometry was performed on the two $z=0.7$ galaxies. A circular 3\arcsec~aperture was used for GSS 044\_6719 and a 5\arcsec~aperture for the larger GSS 044\_6712.  The zeropoints are those discussed in Section~\ref{data}. 
The results are presented in Table~\ref{galaxy_aperture_photometry}.
Our total $K'$ magnitude for GSS 044\_6719 of $18.65\pm{0.02}$ is comparable to seeing-limited photometry from Palomar giving $K=18.52\pm0.04$ \citep{Noeske2006}. Note the similar colors of the two galaxies, especially in $H-K'$. In order to probe the colors of the cores, we repeated the photometry using a smaller circular aperture centered on the brightest pixel in each galaxy in $K'$. The size of 0\farcs4 was chosen to be slightly larger than the NIC3 PSF FHWM, and for $z=0.7$ corresponds to about 3 kpc in the rest frame of the galaxies. Also, for GSS 044\_6719 the edge of this aperture nicely falls in ``trough" between the bright core and the spiral arms in ACS $V$ and $I$. We find that the cores of both galaxies are redder than their total light, especially GSS 044\_6712 in $H-K'$.

As an improvement over our small aperture we employed GIM2D (Marleau \& Simard 1998, Simard et al. 2002) to obtain separated bulge and disk photometry.
The GIM2D code models a galaxy's total flux, $B/T$, bulge and disk
sizes and orientations, and galaxy pixel center from an input image, or simultaneously from multiple images. Sersic bulge plus exponential disk models are convolved with an input PSF and directly compared to data during an optimization with the Metropolis algorithm
\citep{Metropolis1953}. Once the algorithm converges, 99\% confidence intervals are determined via Monte-Carlo sampling of parameter space ($N_{\rm sample}=300$).

A method similar to that described in \cite{Steinbring2004} was applied, adopting a de Vaucouleurs bulge (Sersic index $n=4$) plus exponential disk ($n=1$) model and using the PSFs discussed in Section~\ref{data}. First, we determined the center of the galaxy in the reddest band (NIRC2 $K'$), as that fit was least likely to be affected by dust or by star-forming hotspots. Then, because GIM2D does not allow simultaneous fits in more than two bands, we fit the galaxy structural parameters from the $I$-band ACS image, which has both high resolution and high signal-to-noise.  Structural parameters fit were bulge and disk position angles (PAs) and sizes (disk $R_{\rm scale}$ and bulge $R_{\rm eff}$), disk inclination $i$, and bulge ellipticity $e$. In all bands we kept the center fixed to that found in $K'$, except for our $H$-band fit, in which we allowed the center to wander 0\farcs05 in both the x andy y directions.  Because the intrinsic pixel size of this image is much larger than for the images in other bands, the $H$-band image was difficult to register against the other bands at the sub-pixel level.  With the galaxy center and structural parameters fixed, the following parameters were then allowed to float in all bands: total flux, bulge-to-total flux ratio ($B/T$), and sky background.  

\begin{figure}
\plotone{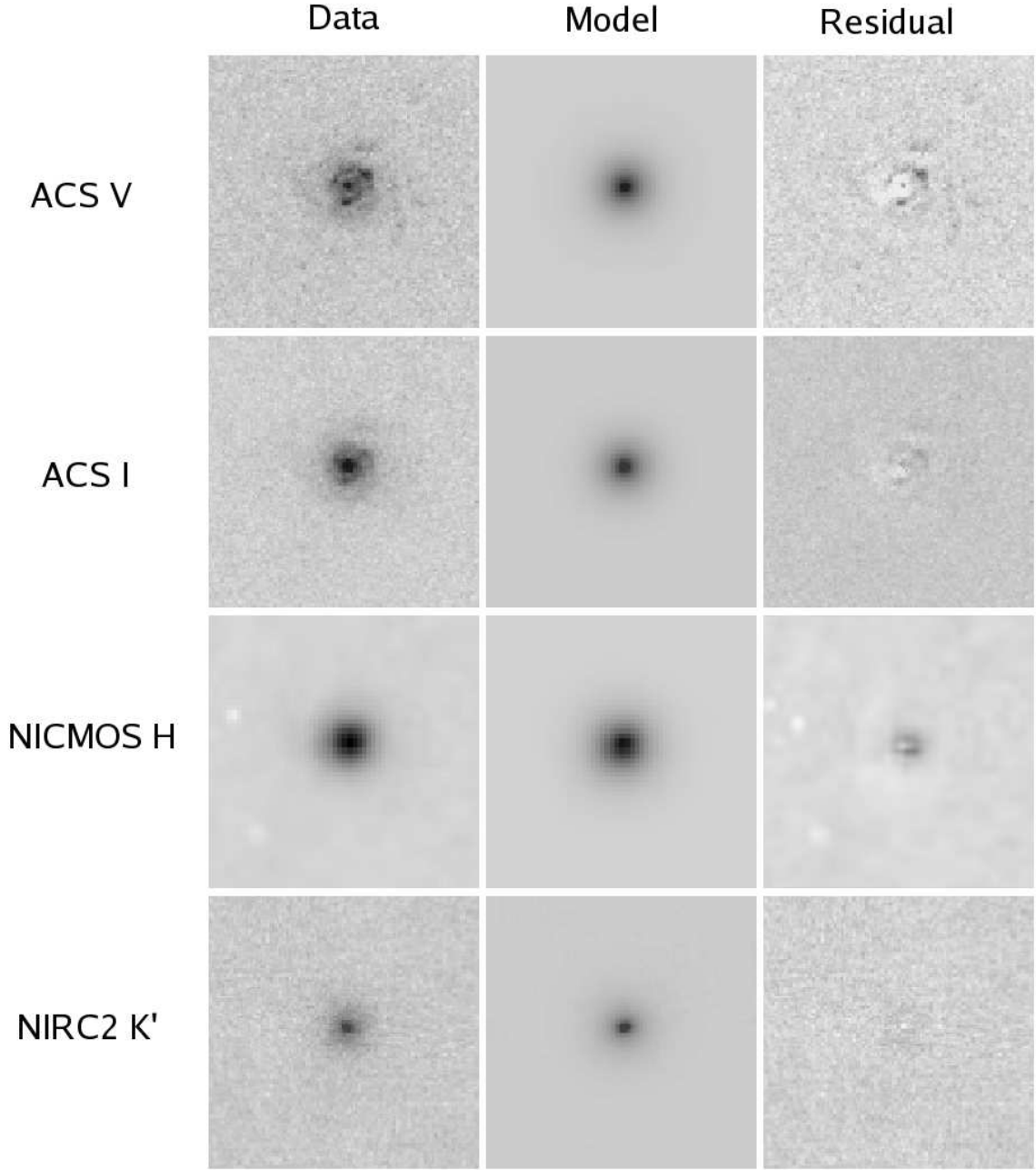}
\caption{Images of the data (left column), models (center column), and residuals (right column) for GSS 044\_6719. The field of view and orientation of each panel is the same as Figure~\ref{figure_panels}. The positive residuals in the ACS images are consistent with clumpy star formation regions in the disk of the galaxy.  Similarily, negative residuals may indicate dust lanes. The simple two-component GIM2D model fits best in $K'$, unconfused by spiral arms and star-formation regions. Note that the NICMOS image has been rebinned to the ACS pixel scale, which makes it appear smoothed.}
\label{figure_models_E}
\end{figure}

\begin{figure}
\plotone{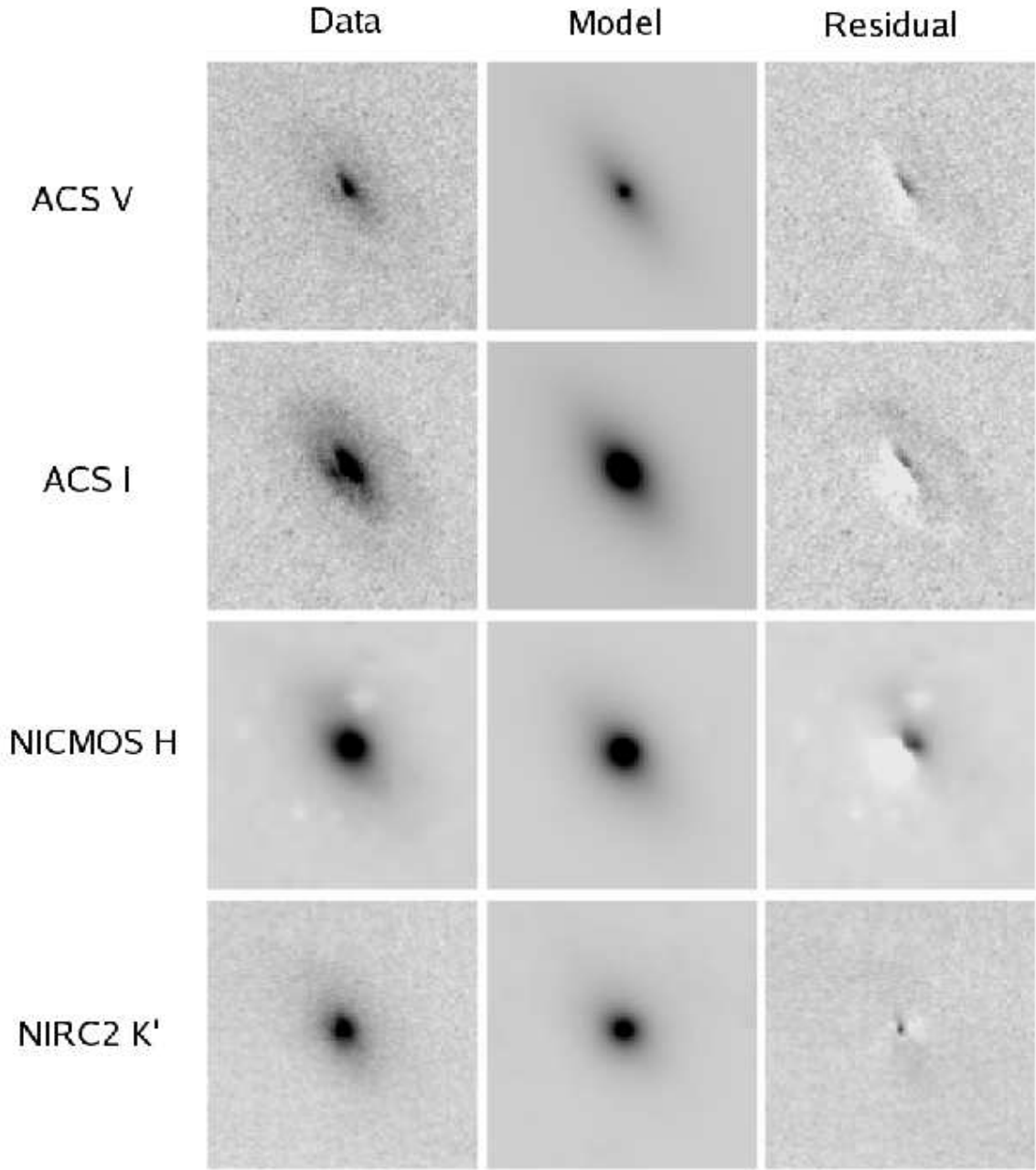}
\caption{Same as Figure~\ref{figure_models_E} except for GSS 044\_6712. The asymmetric core prevented a reliable centering of the model.}
\label{figure_models_C}
\end{figure}

The results are shown for GSS 044\_6719 in Figure~\ref{figure_models_E}. In the left column of panels in this figure, the ACS $V$ and $I$, NICMOS $H$, and NIRC2 $K'$ data are shown.  Best-fit models for each passband are displayed in the center column, and the residual images are to the right.  Note how the positive residuals from modeling the ACS $V$ and $I$ images highlight clumpy star formation regions that are not accounted for in our (smooth) models. Negative residuals in $V$ and $I$ may indicate significant dust in the disk.  For GSS 044\_6712, a dust lane is prominent, and is visible even in $K'$. In fact, the asymmetric core hindered a robust centering of the model, preventing meaningful constraints on what could be a small bulge. Worse results were obtained by centroiding instead on the $I$ image, which is better resolved but less symmetric than in $K'$ (see figure~\ref{figure_models_C}). Masking out the dust lane in the $I$ image did not significantly improve the fit.

The resulting structural parameters for GSS 044\_6719 are shown in Table~\ref{table_parameters}, quoted with 99\% confidence limits. The photometry of the disk and bulge components for GSS 044\_6719 was derived from the total fluxes and the $B/T$. Bulge magnitudes correspond to limits that
have been propagated from the total magnitude measurements and their uncertainties. 
Total, bulge, and disk colors are presented in Table~\ref{galaxy_photometry}; errors again represent 99\% confidence limits.  The redness of the bulge relative to the disk is apparent in both $I-K'$ and $H-K'$, but the former is likely a more reliable measurement due to the better match between NIRC2 and ACS PSFs.

Two tests were carried out on the photometry of GSS 044\_6719 as a check on the colors obtained from GIM2D fitting. The first test was to repeat the GIM2D analysis but instead as a simultaneous fit in ACS $V$ and $I$ only and using PSFs derived from TinyTim. The results of the fit and the correponding photometry are shown in Tables~\ref{table_parameters} and \ref{galaxy_photometry}. That the ACS-only and the combined ACS, NICMOS, and NIRC2 datset agree on an upper limit of $R_{\rm eff}=$ 0\farcs3 is reassuring, as both their PSFs and GIM2D fitting methods are slightly different.
A second test was to carry out aperture photometry in $V$, $I$, and $K'$ in concentric annuli. First, the ACS $I$ image was degraded slightly with a Gaussian filter to match the resolution of the NIRC2 $K'$ image.  Then, the IRAF task ELLIPSE was used to measure photometry in annular apertures with outer radii ranging from 0\farcs1 to 0\farcs8 (in 0\farcs1 increments) in all three bands.  Near-circular apertures were used, with the center, position angle, and ellipticity of each aperture fixed to the best-fit values from GIM2D disk component fits to the ACS $I$ band image of the galaxy.  The resulting aperture photometry is given in Table~\ref{table_aperture_photometry}. Note that only for apertures of outer radius smaller than 0\farcs3 are the $V-I$ and $H-K'$ colors redder than 1.35 and 2.55 magnitudes respectively, the total galaxy colors derived from GIM2D fitting. This confirms that most of the red light is confined to what GIM2D finds as the bulge effective radius. 

\section{Discussion}\label{discussion}

The integrated ACS $I$ magnitude of GSS 044\_6719 is 21.1, which gives an absolute rest-frame $B$ magnitude
of -21.0 using the $K$-corrections of \cite{Simard2002}. For GSS 044\_6712 ($I=20.3$) this is $M_B=-21.4$. These $M_B$, along with a disk scale length $R_{\rm scale}\approx 2.4$ kpc found by GIM2D fitting are similar to those of the three DEEP galaxies we observed in Keck NGS mode reported in \cite{Steinbring2004}. 
These are seemingly normal spirals, near the peak of the luminosity function at $0.6<z<0.8$ as determined in the DEEP2 survey \citep{Willmer2006}. The Keck DEIMOS spectrum from that survey also provides velocity dispersion for GSS 044\_6719 of $\sigma = 24\pm8$ km $s^{-1}$, consistent with a disk-dominated galaxy. From our GIM2D fitting, we can see that while the bulge is weak in the optical (restframe UV), it may comprise 20\% of the $K'$ (restframe~$\sim J$) light. Interestingly this galaxy is detected at $24\mu{\rm m}$ (restframe 14 $\mu$m) with Spitzer \citep{Davis2007}, indicating significant dust may be present.

\begin{figure}
\plotone{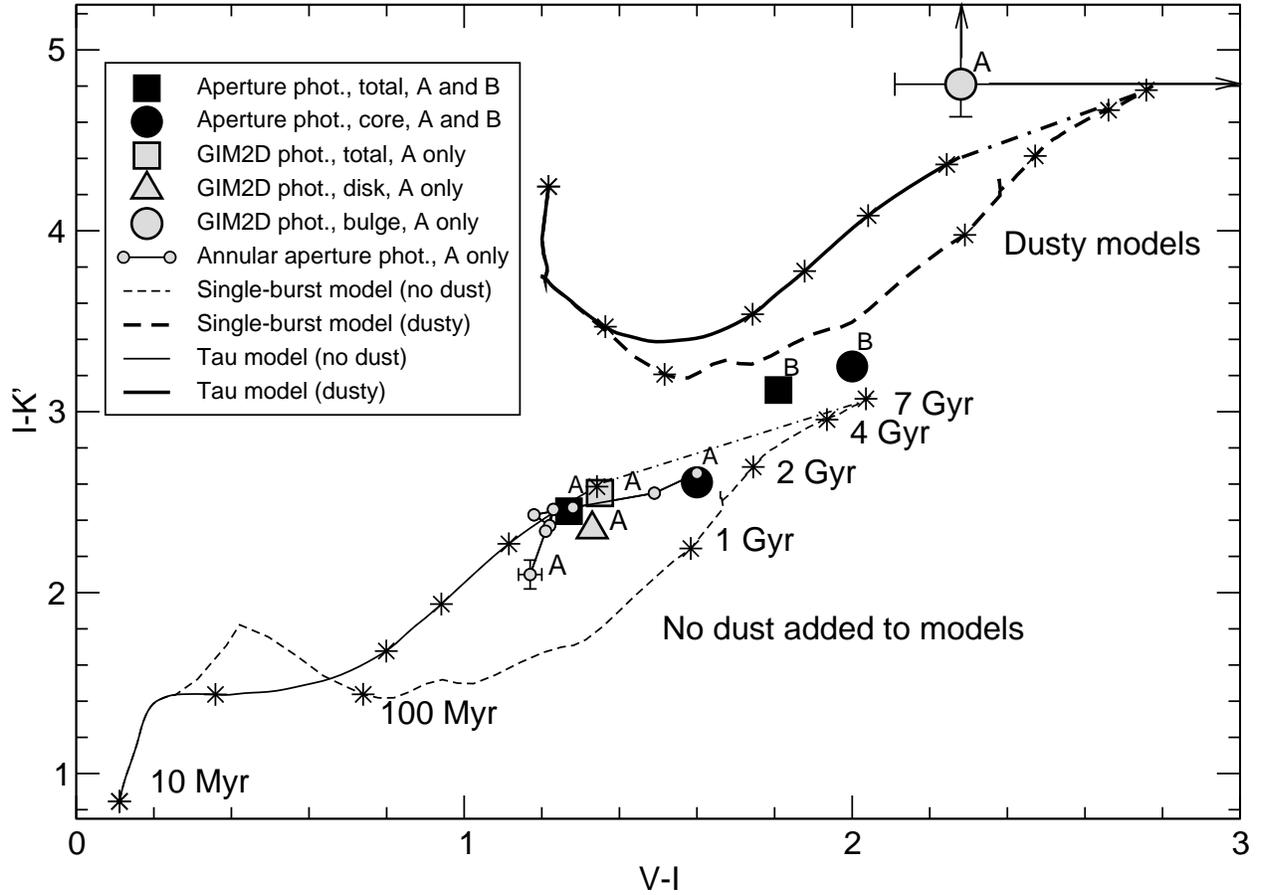}
\caption{Total (black square) and core (black circle) colors from aperture photometry.  Galaxy GSS 044\_6719 is indicated by `A' here; GSS 044\_6712 by `B'. Colors from GIM2D fitting for galaxy A are also shown: total (grey square), disk (grey triangle), and bulge (grey circle). The results from annular aperture photometry of galaxy A are plotted, connected by solid lines. Error bars are shown only for the smallest aperture used, but the others are similar, and are omitted for clarity. Overplotted are the single-burst+passive-evolution model (thin dashed curve) and an exponential $\tau=3$ Gyr model (thin solid curve) without dust, and again with $A_V=10$ magnitudes of extinction (thick upper curves). Stars indicate ages since the onset of star-formation of 10 Myr, 100 Myr, 1 Gyr, 2 Gyr, 4 Gyr, and 7 Gyr; the dot-dashed lines connect the oldest ages in the two star-formation models, delineating the region encompassed by different star-formation histories. The colors of both galaxies can plausibly be explained by the models, with some reddening.  The red $I-K'$ color of the galaxy A bulge would require significant dust to do so.}
\label{figure_bulge_colors}
\end{figure}

In an attempt to understand the very red optical to NIR colors of these galaxies, and especially the bulge of GSS 044\_6719, we used the Galaxev code of \citet{Bruzual2003} to compare
population synthesis models with the data. Galaxev constructs spectra for arbitrary star formation histories by adding together
the single-stellar-population spectra from various ages, weighted appropriately. We produced $V-I$ versus $I-K'$ colors from these
spectra by scaling them according to the \citet{Calzetti1997} dust attenuation curve, shifting them to $z=0.7$, and then convolving this with the appropriate filter bandpasses. We constructed two models, both with solar metalicity ($Z=0.02$). 
The first has an exponential ($\tau=3$ Gyr) star formation rate, and the other passively evolves - that is, there is no active star
formation - after a brief (50 Myr) burst. These were also run assuming a large amount of dust ($A_V=10$), for a total of four models. The results are shown in Figure~\ref{figure_bulge_colors}, for ages of 10 Myr, 100 Myr, 1 Gyr, 2 Gyr, 4 Gyr, and 7 Gyr (the age of the universe at $z=0.7$ for our chosen cosmology) after the onset
of star formation.
The total colors of GSS 044\_6719 from aperture photometry are well fit by an old $\tau$ model with no dust; a value of $\tau>3$ Gyr will just shift the track blueward. The core (0\farcs4 aperture) is redder than the total light, and for a uniform age throughout the galaxy more closely matched by the single-burst+passive model. By adding dust (and keeping $\tau=3$) the same models can also hold for the other galaxy, GSS 044\_6712. However, the bulge of GSS 044\_6719 can only be fit by either a $\tau$ model or a single-burst+passive model after invoking significant dust. One way to reconcile this bulge color with the total light, which seems essentially unreddened, would be a clumpy distribution of dust in the galaxy. We could be seeing an old bulge through a dust clump in a thick disk. The bulge is very small, and so any obscuration would affect its color much more dramatically than the disk. The same might be true for an AGN which, highly obscured, could contribute to these red colors, as well as produce significant mid-IR emission. 

The combination of the high resolution and long wavelength of our NIRC2 AO imaging, along with existing HST imaging, give us an opportunity to measure and interpret the optical to NIR color gradients of the bulge and disk of two galaxies at $z=0.7$.  Aperture photometry shows that the centers of the galaxies are significantly redder than the outer regions.  While we have only explored a few stellar population models in our interpretation of this, we find that the color gradient can be explained by two different star-formation histories for bulge and disk, with similar ages. However, aperture photometry is much more difficult to interpret in terms of galaxy subcomponents, as the light in any given aperture is a combination of light from the galaxy bulge and disk. Furthermore, our GIM2D analysis shows that the bulge of GSS 044\_6719 is extremely small, so only the central aperture we used is likely to contain a significant amount of bulge light.

A much larger sample of galaxies is needed, but it seems we have established a path towards obtaining it.  Employing the Keck laser in our NIRC2 AO observations has resolved the PSF issues that troubled our NGS work and made it particularly difficult to measure and analyze galaxy subcomponent colors.  We have demonstrated a method of dithering the laser guide star that stabilizes the PSF across the field of view during science observations.  Furthermore, our method of interleaving science and PSF calibration frames, and of imaging a crowded stellar field for further calibration, allows us to much better characterize the PSF in our LGS science images than in the NGS images we analyzed in our previous work.  This is important because LGS AO is quickly becoming a standard feature of large telescopes, opening up the sky to deep NIR imaging surveys in archival HST faint galaxy fields. This is the path that the CfAO Treasury Survey is pursuing.  

\acknowledgements

We gratefully acknowledge Matthew Barczys and Shelley Wright for their contributions to CATS. Our thanks go to Jennifer Lotz and Susan Kassin for providing the HST images and Kai Noeske for providing us the 
Palomar $K$ magnitude of GSS 044\_6719.
We appreciate the hard-working Keck staff, especially Observing Assistant Steven Magee and Keck AO team members
Randy Campbell, Antonin Bouchez, David Le Mignant (currently at CfAO), and Peter Wizinowich, who made the huge complexity of laser operations seem (to us) like routine observations.
We acknowledge the great cultural significance of Mauna Kea to native Hawaiians, and express 
gratitude for permission to observe from its summit. Data presented herein were obtained at the 
W. M. Keck Observatory, which is operated as a scientific partnership among the California 
Institute of Technology, the University of California, and the National Aeronautics and Space 
Administration. The Observatory was made possible by the generous financial support of the 
W. M. Keck Foundation. This work was supported by the National Science Foundation Science and 
Technology Center for Adaptive Optics, managed by the University of California at Santa Cruz 
under cooperative agreement No. AST-9876783. AJM appreciates support from the National Science Foundation from grant AST-0302153 through the NSF Astronomy and Astrophysics Postdoctoral Fellows program.

\clearpage

\begin{deluxetable}{ccccc}
\tablecaption{Targets\label{table_targets}}
\tablewidth{0pt}
\tabletypesize{\small}
\tablehead{& &\multicolumn{2}{c}{Coordinates\tablenotemark{1}~~(J2000.0)} &\\
\cline{3-4}
\colhead{Name} &\colhead{IAU} &\colhead{R.A.} &\colhead{Dec.} &\colhead{$z$}}
\startdata
\objectname{GSS 044\_6719} &141809.0+523200 &{14 18 09.05} &{52 32 00.3} &0.67368\\
\objectname{GSS 044\_6712} &141808.8+523207 &{14 18 08.87} &{52 32 07.2} &$0.70^{+0.06}_{-0.02}$~\tablenotemark{2}\\
\nodata &141811.0+523149 &{14 18 11.00} &{52 31 49.8} &1.22828
\enddata
\tablenotetext{1}{Units of right ascension are hours, minutes, and seconds, and units of declination are degrees, arcminutes, and arcseconds.}
\tablenotetext{2}{Photometric redshift. Uncertainties are 99\% confidence limits.}
\end{deluxetable}

\clearpage

\begin{deluxetable}{lcccccccc}
\rotate
\tablecaption{Galaxy Aperture Photometry\tablenotemark{1}\label{galaxy_aperture_photometry}}
\tablewidth{0pt}
\tabletypesize{\small}
\tablehead{\colhead{Target} &\colhead{Aperture} &\colhead{$V$} &\colhead{$I$} &\colhead{$H$} &\colhead{$K'$} &\colhead{$V-I$} &\colhead{$I-K'$} &\colhead{$H-K'$}}
\startdata
GSS 044\_6719 &3\farcs0 &$22.37\pm{0.03}$ &$21.10\pm{0.02}$ &$19.41\pm{0.03}$ &$18.65\pm{0.01}$ &$1.27\pm{0.05}$ &$2.45\pm{0.02}$ &$0.76\pm{0.03}$\\
 &0\farcs4 &$24.86\pm{0.03}$ &$23.26\pm{0.02}$ &$21.68\pm{0.03}$ &$20.65\pm{0.01}$ &$1.60\pm{0.05}$ &$2.61\pm{0.02}$ &$1.03\pm{0.03}$\\
GSS 044\_6712 &5\farcs0 &$22.14\pm{0.03}$ &$20.33\pm{0.02}$ &$18.10\pm{0.03}$ &$17.21\pm{0.01}$ &$1.81\pm{0.05}$ &$3.12\pm{0.02}$ &$0.89\pm{0.04}$\\
 &0\farcs4 &$24.80\pm{0.03}$ &$22.80\pm{0.01}$ &$20.75\pm{0.03}$ &$19.55\pm{0.01}$ &$2.01\pm{0.04}$ &$3.25\pm{0.02}$ &$1.20\pm{0.03}$
\enddata
\tablenotetext{1}{Uncertainties are 1-$\sigma$ limits assuming only Poisson noise.}
\end{deluxetable}

\clearpage

\begin{deluxetable}{lcccccccccc}
\tablecaption{Model Galaxy Parameters for GSS 044\_6719\tablenotemark{1}\label{table_parameters}}
\tablewidth{0pt}
\tabletypesize{\tiny}
\rotate
\tablehead{& & & & & & &\multicolumn{4}{c}{$B/T$}\\
\cline{8-11}
\colhead{Dataset} &\colhead{Disk $R_{\rm scale}$\tablenotemark{2}} &\colhead{$i_{\rm disk}$} &\colhead{${\rm PA}_{\rm disk}$\tablenotemark{3}} &\colhead{Bulge $R_{\rm eff}$\tablenotemark{2}}
&\colhead{$e_{\rm bulge}$} &\colhead{${\rm PA}_{\rm bulge}$\tablenotemark{3}} &\colhead{$V$} &\colhead{$I$} &\colhead{$H$} &\colhead{$K'$}\\
&\colhead{(arcsec)} &\colhead{(deg)} & &\colhead{(arcsec)} & & & & & &}
\startdata
ACS $V$ and $I$, NICMOS $H$, \& NIRC2 $K'$\tablenotemark{4} &$0.33^{+0.01}_{-0.01}$ &$13.7^{+4.1}_{-6.5}$ &$-62.9^{+9.3}_{-22.1}$ &$0.01^{+0.29}_{-0.01}$ &$0.20^{+0.17}_{-0.20}$ &$41.5^{+25.6}_{-136.8}$ &$0.01^{+0.01}_{-0.01}$ &$0.02^{+0.08}_{-0.01}$ &$0.05^{+0.01}_{-0.01}$ &$0.19^{+0.02}_{-0.01}$\\
ACS $V$ and $I$\tablenotemark{5} &$0.39^{+0.01}_{-0.01}$ &$9.0^{+3.4}_{-5.8}$ &$95.4^{+18.2}_{-5.8}$ &$0.21^{+0.08}_{-0.09}$ &$0.27^{+0.11}_{-0.03}$ &$80.5^{+12.8}_{-30.7}$ &$0.01^{+0.01}_{-0.01}$ &$0.07^{+0.02}_{-0.02}$ &\nodata &\nodata
\enddata
\tablenotetext{1}{Structural parameters are described in Section 3.  Uncertainties are 99\% confidence limits.}
\tablenotetext{2}{For $z=0.7$, 1 kpc corresponds to 0\farcs14 on the sky assuming $H_0=70$ km ${\rm s}^{-1}$ Mpc, $\Omega_{\rm m}=0.3$, and $\Omega_{\Lambda}=0.7$.}
\tablenotetext{3}{Counter-clockwise from north.}
\tablenotetext{4}{Independent fit in each band, with a star used as PSF.}
\tablenotetext{5}{Simultaneous fit in $V$ and $I$ using a TinyTim model PSF.}
\end{deluxetable}

\clearpage

\begin{deluxetable}{lccccccc}
\rotate
\tablecaption{Photometry of GSS 044\_6719 from GIM2D\tablenotemark{1}\label{galaxy_photometry}}
\tablewidth{0pt}
\tabletypesize{\small}
\tablehead{\colhead{Component} &\colhead{$V$} &\colhead{$I$} &\colhead{$H$} &\colhead{$K'$} &\colhead{$V-I$} &\colhead{$I-K'$} &\colhead{$H-K'$}}
\startdata
\sidehead{ACS, NICMOS, \& NIRC2 data fit with $I$-band parameters}
Total &$22.25^{+0.02}_{-0.02}$ &$20.90^{+0.01}_{-0.01}$ &$19.18^{+0.04}_{-0.03}$ &$18.35^{+0.03}_{-0.02}$ &$1.35^{+0.02}_{-0.02}$ &$2.55^{+0.03}_{-0.03}$ &$0.83^{+0.04}_{-0.05}$\\
Disk  &$22.26^{+0.02}_{-0.02}$ &$20.93^{+0.10}_{-0.01}$ &$19.23^{+0.04}_{-0.03}$ &$18.58^{+0.04}_{-0.03}$ &$1.33^{+0.02}_{-0.10}$ &$2.35^{+0.03}_{-0.10}$ &$0.66^{+0.04}_{-0.06}$\\
Bulge &$27.25^{+0.02}_{-0.02}$ &$24.97^{+0.17}_{-1.65}$ &$22.42^{+0.05}_{-0.09}$ &$20.16^{+0.07}_{-0.10}$ &$2.28^{+1.65}_{-0.17}$ &$4.81^{+1.65}_{-0.18}$ &$2.26^{+0.13}_{-0.09}$\\
\sidehead{ACS data only}
Total &$22.23^{+0.02}_{-0.03}$ &$20.99^{+0.02}_{-0.03}$ &\nodata &\nodata &$1.25^{+0.02}_{-0.03}$ &\nodata &\nodata\\
Disk  &$22.24^{+0.02}_{-0.03}$ &$21.00^{+0.02}_{-0.03}$ &\nodata &\nodata &$1.18^{+0.05}_{-0.04}$ &\nodata &\nodata\\
Bulge &$27.82^{+2.68}_{-0.98}$ &$26.57^{+2.68}_{-0.98}$ &\nodata &\nodata &$3.91^{+0.99}_{-2.56}$ &\nodata &\nodata 
\enddata
\tablenotetext{1}{Uncertainties are 99\% confidence limits.}
\end{deluxetable}

\clearpage

\begin{deluxetable}{cccccc}
\tablecaption{Annular Aperture Photometry of GSS 044\_6719 on ACS and NIRC2 Images\tablenotemark{1}\label{table_aperture_photometry}}
\tablewidth{0pt}
\tabletypesize{\small}
\tablehead{\colhead{Aperture\tablenotemark{2}} &\colhead{$V$} &\colhead{$I$} &\colhead{$K'$} &\colhead{$V-I$} &\colhead{$I-K'$}}
\startdata
0\farcs0 to 0\farcs1 &$26.15_{-0.02}^{+0.02}$ &$24.55_{-0.02}^{+0.02}$ &$21.89_{-0.05}^{+0.05}$ &$1.60_{-0.04}^{+0.04}$ &$2.66_{-0.07}^{+0.07}$\\
0\farcs1 to 0\farcs2 &$25.24_{-0.02}^{+0.02}$ &$23.79_{-0.01}^{+0.01}$ &$21.24_{-0.03}^{+0.03}$ &$1.49_{-0.03}^{+0.03}$ &$2.55_{-0.04}^{+0.04}$\\
0\farcs2 to 0\farcs3 &$24.78_{-0.02}^{+0.02}$ &$23.50_{-0.01}^{+0.01}$ &$21.03_{-0.02}^{+0.02}$ &$1.28_{-0.03}^{+0.03}$ &$2.47_{-0.03}^{+0.03}$\\
0\farcs3 to 0\farcs4 &$24.57_{-0.04}^{+0.04}$ &$23.39_{-0.02}^{+0.02}$ &$20.96_{-0.02}^{+0.02}$ &$1.18_{-0.06}^{+0.06}$ &$2.43_{-0.04}^{+0.04}$\\
0\farcs4 to 0\farcs5 &$24.69_{-0.04}^{+0.05}$ &$23.47_{-0.02}^{+0.02}$ &$21.10_{-0.03}^{+0.03}$ &$1.22_{-0.06}^{+0.07}$ &$2.37_{-0.05}^{+0.05}$\\
0\farcs5 to 0\farcs6 &$24.89_{-0.03}^{+0.03}$ &$23.66_{-0.02}^{+0.02}$ &$21.20_{-0.03}^{+0.04}$ &$1.23_{-0.05}^{+0.05}$ &$2.46_{-0.05}^{+0.06}$\\
0\farcs6 to 0\farcs7 &$25.00_{-0.03}^{+0.03}$ &$23.79_{-0.02}^{+0.02}$ &$21.45_{-0.05}^{+0.06}$ &$1.21_{-0.05}^{+0.05}$ &$2.34_{-0.07}^{+0.08}$\\
0\farcs7 to 0\farcs8 &$25.07_{-0.02}^{+0.02}$ &$23.90_{-0.01}^{+0.01}$ &$21.80_{-0.07}^{+0.07}$ &$1.17_{-0.03}^{+0.03}$ &$2.10_{-0.08}^{+0.08}$
\enddata
\tablenotetext{1}{Uncertainties are 1-$\sigma$ limits assuming only Poission noise.}
\tablenotetext{2}{The inner and outer radii of the annular aperture are given.}
\end{deluxetable}

\clearpage

\clearpage

\clearpage

\clearpage

\clearpage

\clearpage

\clearpage

\clearpage

\clearpage

\end{document}